\documentstyle[12pt]{article}
\def\singlespace {\smallskipamount=3.75pt plus1pt minus1pt
                  \medskipamount=7.5pt plus2pt minus2pt
                  \bigskipamount=15pt plus4pt minus4pt
                  \normalbaselineskip=15pt plus0pt minus0pt
                  \normallineskip=1pt
                  \normallineskiplimit=0pt
                  \jot=3.75pt
                  {\def\smallskip {\vskip\smallskipamount}}
                  {\def\medskip   {\vskip\medskipamount}}
                  {\def\bigskip   {\vskip\bigskipamount}}
                  {\setbox\strutbox=\hbox{\vrule
                    height10.5pt depth4.5pt width 0pt}}
                  \parskip 7.5pt
                  \normalbaselines}
\def\middlespace {\smallskipamount=5.825pt plus1.5pt minus1.5pt
                  \medskipamount=11.25pt plus3pt minus3pt
                  \bigskipamount=22.5pt plus6pt minus6pt
                  \normalbaselineskip=22.5pt plus0pt minus0pt
                  \normallineskip=1pt
                  \normallineskiplimit=0pt
                  \jot=5.825pt
                  {\def\smallskip {\vskip\smallskipamount}}
                  {\def\medskip   {\vskip\medskipamount}}
                  {\def\bigskip   {\vskip\bigskipamount}}
                  {\setbox\strutbox=\hbox{\vrule
                    height15.75pt depth6.75pt width 0pt}}
                  \parskip 5.25pt
                  \normalbaselines}
\def\dblspc {\smallskipamount=7.5pt plus2pt minus2pt
                  \medskipamount=15pt plus4pt minus4pt
                  \bigskipamount=30pt plus8pt minus8pt
                  \normalbaselineskip=30pt plus0pt minus0pt
                  \normallineskip=2pt
                  \normallineskiplimit=0pt
                  \jot=7.5pt
                  {\def\smallskip {\vskip\smallskipamount}}
                  {\def\medskip   {\vskip\medskipamount}}
                  {\def\bigskip   {\vskip\bigskipamount}}
                  {\setbox\strutbox=\hbox{\vrule
                    height21.0pt depth9.0pt width 0pt}}
                  \parskip 12.0pt
                  \normalbaselines}
\def\al{\alpha}
\def\nb{\nabla}
\def\eps{\epsilon}
\def\sch{Schwarzschild }

\def\la {\lambda }

\def\be{\begin{equation}}

\def\j-{\J_-}

\def\ee{\end{equation}}

\def\bearr{\begin{eqnarray}}
\def\bearrs{\begin{eqnarray*}}
\def\eearr{\end{eqnarray}}
\def\eearrs{\end{eqnarray*}}
\def\barr{\begin{array}}
\def\earr{\end{array}}
\def\p{\partial}

\def\o{\omega}

\def\non\non{\nonumber}
\def\nn8{\nonumber\\[15pt]}
\def\l{\left}
\def\r{\right}
\def\un{\underline}

\def\f{\frac}

\def\dis{\displaystyle}

\begin{document}
\thispagestyle{empty}
\middlespace
\begin{center}
{\large{\bf  Photon propagation in Einstein and\\[6pt] Higher
Derivative Gravity}}\\[20pt]
Subhendra Mohanty and A.R. Prasanna\\
Physical Research Laboratory\\
Ahmedabad 380 009, India\\[30pt]
\un{Abstract}\\
\end{center}

We derive the wave equation obeyed by electromagnetic fields in
curved spacetime.  We find that there are Riemann and Ricci
curvature coupling terms to the photon polarisation which result
in a polarisation dependent deviation of the photon trajectories
from null geodesics.  Photons are found to have an effective
mass in an external gravitational field and their velocity in an
inertial frame is in general less than $c$.  A physically relevant
consequence of the analysis is that the
curvature corrections to the propagation of electromagnetic
radiation (in a homogenous and isotropic spacetime) keep the velocities
subluminal provided the strong
energy condition is satisfied.  We further show that the claims
of superluminal velocities in higher derivative gravity theories
are erroneous and arise due to the
neglect of Riemann and Ricci coupling terms in the wave
equation, of Einstein gravity. \\

\newpage

A standard result of Einstein's gravity is that the trajectories
of all massless particles are null geodesics.  A question worth
examining is whether there is a deviation from the null
geodesics if the particles have a spin i.e. due to the
interaction of spin with the Riemann and Ricci curvatures of the
gravitational fields.  In this paper we have studied this
question for the case of photon propagation in a curved
background.  Starting with the action
$\sqrt{-g} F_{\mu\nu}\; F^{\mu\nu}$ for electromagnetic fields
in a 
gravitational field, we derive the wave equation
for electromagnetic field tensor $F_{\mu\nu}$, which turns out
to be of the form (Eddington [1] , Noonan 
[2])
\be
\nb^\mu \nb_\mu F_{\nu \la} + R_{\rho\mu \nu \la} F^{\rho \mu} -
R^\rho_{\;\;\la} F_{\nu \rho} + R^\rho_{\;\;\nu} F_{\la \rho} = 0
\ee
We see that the photon propagation depends upon the coupling
between the Riemann and Ricci curvatures and the photon
polarisation. This leads to a deviation of the photon
trajectories from the null geodesic by amounts proportional to
the Riemann and Ricci curvatures.  The photon trajectories in
the geometrical optics limit are described by the following
generalisation of the geodesic equation
\be
\f{d^2X^a}{ds^2} + \Gamma^\al_{\beta\gamma} \f{dX^\beta}{ds}
\f{dX^\gamma}{ds} = \f{1}{2} \nb^\alpha \l[ R^{\rho \mu}_{\;\;\;\;\;\nu\la}
f_{\rho\mu} 
+ R^\rho_{\;\;\la} f_{\nu \rho} - R^\rho_{\;\;\nu} f_{\la \rho}\r] 
 \times \f{f^{\nu\la}}{\mid f^2\mid}
\ee
The nonzero right hand side of the modified geodesic equation
(2) implies that there is a polarisation dependence in the
gravitational red
shift,  bending, and  Shapiro time delay of light even in classical
Einsteins gravity.

A curious phenomena discovered by Drummond and Hathrell [4] is
that in higher derivative gravity which arises by QED radiative
corrections, the photon velocity in local inertial frame can
exceed the velocity of light in the Minkowski space.  Due to the
coupling of the electromagnetic fields with the Riemann and
Ricci tensors in the Lagrangian, it was claimed that the photon
velocity in the Schwarzschild, Robertson-
Walker, gravitational
wave and deSitter backgrounds is larger than the flat space
velocity $c$.  This result was extended by Daniels and Shore to
charged [5] and rotating blackholes [6] with the same
conclusions.  Latorre et al [7] have shown a universal relation
between the velocity shift of photons to the energy density
which generates the background metric, and Shore [8] has related
the velocity shift to the coefficients of conformal anomaly.
Lafrance and Myers [9] interpret these results as the breakdown
of the Equivalence principle in higher derivative gravity and
Dolgov and Khriptovich [10,11] derive this result from field
theoretic dispersion relations. Finally Mende [12] has proposed
this effect as a test for string theories of quantum gravity.

In this paper we show that claims of superluminal photon velocity are due
to the neglect
of the Riemann coupling terms in the wave equation which arises
from the minimal $F_{\mu\nu} F^{\mu\nu}$ Lagrangian.  We find that
in Einstein's gravity the photon velocity in a \sch blackhole
metric 
and in the Friedman-Robertson-Walker metric is {\it less} than $c$. In
other words the photon trajectories are always inside the null cone.
{\it We find that for the photon trajectories not to go out of the null cone  
the background matter should
satisfy the strong energy condition} $\rho \ge 3p $.  
This provides  us with  an  answer to the question raised by
Zeldovich and Novikov [13] - What law of physics would be violated
if the strong energy condition is not satisfied ? Our answer is that
special relativity in the free fall inertial- frame demands that the
strong energy
condition be satisfied. 

We also derive the wave in higher derivative gravity 
and show that
Riemann coupling terms in the lagrangian of the higher derivative gravity
are
always smaller in magnitude than the Riemann term that already
exists in Einstein's gravity.  This analysis shows that the
photon velocity does not exceeds $c$, even  
 by the inclusion of the radiative correction higher derivative terms in
the Lagrangian contrary to the claims [4-12]. The absence of superluminal
propagation is not dependent on the eikonal ansatz but is a consequence
of the fact that the wave-equation in for the electromagnetic fields
 in Einstein
and  higher derivative gravity  is {\it hyperbolic}. Consequently the
field solution at a given point can only depend upon sources {\it inside}
the past null-cone. In references [4-12] the vector potential solution is
assumed to be of the eikonal form, and it is shown that the vector
potential propagates outside the null-cone.
The eikonal  approximation for the vector potential does not
always correctly describe the propagation of electromagnetic
waves since the vector potential is not gauge invariant and it can
be non-zero even in the acausal regions where the
electromagnetic field is zero. 

The interaction of electromagnetic fields with gravity is given
by the action
\be
S = \int d^4x \sqrt{-g} F_{\mu\nu} F^{\mu\nu}
\ee
(We use the convention $c=1$, signature $-2$ and Greek letters
denote spacetime indices $0-3$). From
(3) we obtain the equations of motion
\be
\nb^\mu F_{\mu\nu} = 0
\ee
Equation (4) with the Bianchi identity
\be
\nb_\mu F_{\nu \la} + \nb_\nu F_{\la \mu} + \nb_\la F_{\mu \nu}
= 0
\ee
gives the Maxwell's equation in curved background. Operating on
equation (4) by the $\nb_\la$ and using the Bianchi identity
(5) the wave equation may be obtained as
\be
\barr{lll}
\nb^\mu \nb_\mu F_{\nu \la} &+& \nb_\la \l( \nb^\mu F_{\nu\mu}
\r) + \l[ \nb^\mu , \nb_\nu \r] F_{\la \mu}\\[8pt]
&-& \l[ \nb_\la , \nb^\mu \r] F_{\mu\nu} = 0
\earr
\ee
The second term vanishes owing to (4). Using the identity for
the commutator of covariant derivatives:
\be
\l[ \nb^\mu , \nb_\nu \r] F_{\al \mu} = R_{\rho \al \nu \mu}
F^{\rho \mu} + R_{\rho \la} F_{\nu}^\rho
\ee
and the circular identity
\be
R_{\rho \la \nu \mu} + R_{\rho \nu \mu \la} + R_{\rho \mu \la
\nu} = 0
\ee
the wave equation (6) reduces to the form 
\be
\nb^\mu \nb_\mu F_{\nu \la} + R_{\rho \mu \nu \la} F^{\rho \mu}
+ R^\rho_{\;\;\la} F_{\nu \rho} - R^\rho_{\;\;\nu} F_{\la \rho} = 0
\ee
The Riemann and Ricci curvature coupling terms to the photon
polarisation (or spin) give rise to the polarisation dependent
deviation of photon orbits from null geodesics.

Photon trajectories are described by the eikonal solutions of
the wave equation (9) of the form
\be
F_{\mu\nu} = e^{iS(x)}f_{\mu\nu}
\ee
where the phase $S$ varies more rapidly in spacetime than the
amplitude $f_{\mu\nu}$.  The wavenumber of the photon
trajectories is given by the gradient of the phase
\be
K_\mu = \nb_\mu S
\ee
In the geometrical optics approximation where
\be
\nb_\mu F_{\al \beta} = i K_\mu F_{\al \beta}
\ee
the wave equation (9) can be written in the form
\be
- K^\mu K_\mu f_{\nu \la} + 
R^{\rho \mu}_{\;\;\;\;\;\nu\la} f_{\rho \mu} - 
R^\rho_{\;\;\la} f_{\nu \rho} + R^\rho_{\;\;\nu} f_{\la \rho} = 0
\ee
The
dispersion relation may be written as
\be
K^2 = \l( R^{\rho\mu}_{\;\;\;\;\;\nu_\la} f_{\rho\mu} - R^\rho_{\;\;\la}
f_{\nu\rho} + R^\rho_{\;\;\nu} f_{\la \rho} \r) \f{f^{\nu\la}}{\l(
f_{\al \beta} f^{\al \beta} \r)} 
\ee
Operating on (14) by $\nb^\al$, the L.H.S. is
\be
\nb^\al K^2 = 2 K^\mu \nb_\mu K^\al
\ee
where we have used the identity $\nb^\al K_\mu = \nb_\mu K^\al$
which follows from the definition of $K_\mu$ as a gradient.
Light rays are defined as the integral curves of the wave vector
$K_\mu$ i.e. the curves $x_\mu(s)$ for which $\f{dX_\mu}{ds} =
K_\mu$.  Substituting $K_\mu = dX_\mu/ds$ in (15) we have for the L.H.S.
\be
\nb^\al K^2 = 2 \f{dX^\mu}{ds} \nb_\mu \l( \f{dX^\al}{ds} \r)
= 2 \l( \f{d^2X^\al}{ds^2} + \Gamma^\al_{\mu\nu}
\f{dX^\nu}{ds} \f{dX^\mu}{ds} \r)
\ee
Using (14), (15) and (16) we obtain the modified geodesic equation
for photon trajectories
\be
\f{d^2X^\al}{ds^2} + \Gamma^\al_{\beta\gamma} \f{dX^\beta}{ds}
\f{dX^\gamma}{ds} = \f{1}{2} \nabla^\alpha \l( R^{\rho
\mu}_{\;\;\;\;\;\nu\la} \cdot 
f_{\rho \mu} - R^\rho_{\;\;\la} f_{\nu\rho} + R^\rho_{\;\;\nu}
f_{\la \rho} \r)
\times \f{f^{\nu\la}}{\l( f^{\al \beta} f_{\al \beta} \r)} 
\ee

To obtain the photon velocity in the curved space, we can use
the dispersion relation (13) directly.  Of the six components of
$F_{\mu\nu}$ in equation (13), only three are independent owing
to the Bianchi identity (5). Choosing the components of the
electric field vector $E_i = f_{oi}$ as the independent
components we have from the Bianchi identity (5)
\be
K_o f_{ij} + K_i f_{jo} + K_j f_{oi} = 0
\ee
Using (18) to substitute for $f_{ij}$ in terms of the electric
field components $f_{jo}$ and $f_{oi}$ in the wave equation (13)
we obtain
\be
\barr{lll}
\l[ \l( + K^\mu K_\mu + R^o_{\;\;o} + R^\ell{\;\;_o} \f{K_\ell}{K_o} \r)
\delta^j_i \right. & +&  \l( -2 R^{oj}_{\;\;\;\;\;oi} + 4
R^{\ell j}_{\;\;\;\;\;oi}  
\f{K_\ell}{K_o} + R^j_{\;\;i} \right.\\[8pt]
&-&  \left.\f{1}{K_o} R^j_{\;\;o} k_i \r) f_{oj} = 0
\earr
\ee
which is the wave equation obeyed by the three components of the
electric field vector (in (19) Latin indices are over three
spacelike components and repeated indices are summed over).  To
simplify the notation we write (15) as 
\be
\l( K^2 \delta^j_i + \eps^j_{i} \r) f_{oj} = 0
\ee
where
\be
\eps^j_i \equiv \l( R^o_{\;\;o} + R^\ell_{\;\;o} \f{K_\ell}{K_o} \r)
\delta^j_i + \l( -2 R^{oj}_{\;\;\;\;oi} + 4R^{\ell
j}_{\;\;\;\;oi} \; \f{K_\ell}{K_o} + 
R^j_{\;\;i} 
- \f{1}{K_o} R^j_{\;\;o} K_i \r)
\ee
Of the three components of the electric field vector in the
electromagnetic waves, only the transverse waves are observable
in an asymptotically flat space.  To obtain the equation for the
transverse fields we first diagonalise the matrix $\eps^j_i$ to
give the equation for the three normal modes:
\be
\l( K^2 + \eps_i \r) f_{oi} = 0 \; \; (i = 1,2,3)
\ee
where $\eps_i$ are the eigenvalues of the $\eps^i_j$ matrix.
The three equations (22) can be separated into the equation for
the transverse fields
\be
f^{(T)}_{oj} \equiv \l( \delta^i_j - n^in_j \r) f_{oi} 
\ee
and the longitudinal fields
\be
f^{(L)}_{oj} \equiv n^i n_j f_{oi}
\ee
where $n^i \equiv \f{K_i}{\mid \vec{K}\mid}$ are the components of
the unit vector along the direction of propagation.  The wave
equation for the transverse photon is
\be
\l[ K^2 \l( \delta^j_{i} - n^j n_i \r) + \eps_i \l( \delta^j_{i} -
n^j n_i \r) \r] f_{oj} = 0
\ee
and for the longitudinal photon the wave equation is
\be
\l( K^2 + \eps_i \r) \l( n^j n_i \r) f_{oj} = 0
\ee

\noindent
\un{Expanding Universe (Friedmann-Robertson-Walker Metric)}:

For wave propagation in a general homogeneous, isotropic
universe described by the line element  $ds^2$ = $+ dt^2$ $- R^2(t) \l(
dr^2 + r^2 d\theta^2 + \sin^2 
\theta d\phi^2 \r)$, 
the non-zero components of the $\eps^i_j$ matrix (21) are
\be
\eps^1_1 = \eps^2_2 = \eps^3_3 = -2 \l( \f{\ddot{R}}{R} +
\f{\dot{R}^2}{R^2} \r) = - \f{8\pi G}{3} \l( \rho -3 p \r)
\ee
where $R$ is the scale factor, and $p$ and $\rho$ are the
pressure and energy density respectively.

The dispersion relation for the transverse photons is
\be
\o^2 - k^2_i = \f{8\pi G}{3} \l( \rho - 3p \r)
\ee
and the photon velocity is given by
\be
v^{(T)}_i = \f{\p \o}{\p k_i} = \f{1}{\l( 1 + \f{8\pi G}{3
k^2_i} \l( \rho -3p \r) \r)^{1/2}} .
\ee
Therefore we find that the photons are not superluminal as long as
the 
strong energy conditon $\rho \ge 3p $ is satisfied. In the radiation
dominated era when $\rho=3p$ the photon velocity is 1.
One may turn this argunment around and use the axiom from special
relativity that photons of any polarisation cannot exceed the flat space
velocity $c$ when measured in an inertial frame,
to show that the strong energy condition $\rho - 3p \ge 0$ must be
obeyed for any form of
matter. Consider a space
time region filled with some matter with equation of state $\rho = q p$. 
The $\eps^i_j$ matrix in that region of spacetime is given by $\eps^i_j
=  - \delta^i_j   (8\pi G \rho /3)  \l( 1 -(3/q)  \r) $ (assuming the
matter
distribution to be homogenous and isotropic though time dependent). The
velocity of photons through that medium is given by $v= c \l( 1 +
(8\pi G \rho
/3k^2_i) \l( 1 -q/3 \r) \r)^{- 1/2}$. Therefore if we assume that
the 
photon velocity in an inertial frame does not exceed that flat space
velocity $c$, then the  equation of state of any form of matter $\rho = q
p$ 
must obey the
strong energy condition $q \ge 3$.

\noindent
\un{Non-Rotating Black Hole Spacetimes (\sch Metric)}:

In the gravitational field of non-rotating blackhole, the
non-zero components of the $\eps^i_j$ matrix (21) are
\be
\eps^1_1 = 4 \f{GM}{r^3}\; , \; \eps^2_2 = \eps^3_3 = -2 \f{GM}{r^3}
\ee
Considering radial trajectories with $\vec{n} = \l( \hat{r},
\hat{\theta}, \hat{\phi} \r) = (1,0,0)$, equation (25) yields
the wave equations for the transverse fields $E_2, E_3$:
\be
\l( \barr{cc}K^2-2GM/r^3&0\\0&K^2-2GM/r^3 \earr \r) \l(
\barr{c}E_2\\E_3 \earr \r) = 0
\ee
The dispersion relation yields
\be
\o^2 - k^2_1 = 2 \f{GM}{r^3}
\ee
The velocity of propagation of the transverse photon is
\be
v^{(t)}_1 = \f{\p \o}{\p k_1} = \f{1}{\l( 1 + \f{1}{k^2_1} \l(
\f{2GM}{r^3} \r) \r)^{1/2}} < 1
\ee
The photon velocity is subluminal and the transverse photons
have an effective mass
\be
m_\gamma = \l( \f{2GM}{r^3} \r)^{1/2}
\ee
For tangential trajectories, $\vec{n} = (0,1,0)$ and the
equations for the transverse fields $E_1$ and $E_{3}$ yield:
\be
\l( \barr{cc}K^2 + 4GM/r^3&0\\0&K^2 - 2GM/r^3 \earr \r) \l(
\barr{c} E_1\\E_3 \earr \r) = 0
\ee
The dispersion relation is $\l( K^2 + \f{4GM}{r^3} \r) \l( K^2 -
\f{2GM}{r^3} \r) = 0$ and the root corresponding to the propagating
mode is
\be
K^2 = \o^2 - k^2_2 = \f{2GM}{r^3}
\ee
and the velocity of wave propagation is given by
\be
v^{(T)} = \f{\p \o}{\p k_2} = \f{1}{\l( 1 + \f{1}{k_1^2} \l(
\f{2GM}{r^3} \r) \r)^{1/2}} < 1
\ee

In the coordinate frame the dispersion relation obeyed by $K_\mu
= \l( \o , k_r k_\theta , k_\rho \r)$ obtained from
the wave equation (25), for a radial trajectory is given by
\be
\l( 1 - \f{2GM}{r} \r)^{-1} \o^2 - \l( 1 - \f{2GM}{r} \r) k^2_r
= \f{2GM}{r^3}
\ee
Since $\o$ is a constant of motion (the metric being stationary),
the wavenumber $k_r$ changes with $r$.  The
redshift of the wavelength $\la = \l( g_{rr} \r)^{1/2} /k_r$ is given by
the
expression
\be
\f{\la\l( r_2 \r)}{\la \l( r_1 \r)} =  \l( \f{\dis  1 - 2GM/r_2}{\dis  1 - 2GM/r_1
}\r)^{1/2}
\l\{ \f{\dis 1 - \f{2GM}{\o^2r^3_1} \l( 1 - \f{2GM}{r_1}
\r)}{\dis 1 - \f{2GM}{\o^2r_2^3} \l( 1 - \f{2GM}{r_2} \r)} \r\}^{1/2}
\ee 
The extra factor in the curly brackets is the correction to the
standard redshift formula due to the coupling of the Riemann
curvature to the wave vector.  The correction is significant when the photon
wavelength is comparable to the Riemann curvature.

\noindent
\un{Higher Derivative Gravity}:

Drummond and Hathrell [3] obtained the higher derivative
couplings which arise from the QED loop corrections to the
graviton-photon vertex.  The effective Lagrangian valid in the
frequency range $\o^2 < \alpha / m^2_e$ is given by
\be
\barr{lll}
{\cal L} &=& - \f{1}{4} F_{\mu\nu} + a R F_{\mu\nu} F^{\mu\nu} +
b R_{\mu\nu} F^{\mu \sigma} F^{\nu}_\sigma\\[8pt]
&& + c R_{\mu\nu\sigma\tau} F^{\mu\nu} F^{\sigma \tau}
\earr
\ee
with the coefficients $a = - \f{5}{720} \f{\alpha}{\pi m^2_e}, b
= \f{26}{620} \f{\alpha}{\pi m^2_e}$, $c = - \f{2}{720}
\f{\alpha}{\pi m^2_e}$ (where $\al$ is the fine structure
constant and $m_e$ the electron mass).  The equations of motion
from (40) are given by
\be
\nb_\mu F^{\mu\nu} + X^\nu = 0
\ee
where
\be
\barr{lll}
X^\nu&=& a \l( F^{\mu\nu} \nb_\mu R + R \nb_\mu F^{\mu\nu} \r)\\[8pt]
&&+ 2b \l( F^{\sigma \nu} \nb_\mu R^\mu_{\;\;\sigma} 
- F^{\sigma \mu} \nb_\mu R^\nu_{\;\;\sigma} - R^\nu_{\;\;\sigma}
\nb_\mu F^{\sigma \mu} \r)\\[8pt]
&&+ 4c \l( F^{\sigma \tau}\nb_\mu R^{\mu \nu}_{\;\;\;\;\sigma\tau} +
R^{\mu\nu}_{\;\;\;\;\sigma\tau} \nb_\mu F^{\sigma\tau} \r) = 0
\earr
\ee
Operating on equation (46) by $\nb^\la$ and using the Bianchi
identity (5) the commutator identity (7) and the circular
identity (7), we obtain the wave equation in higher derivative
gravity: 
\be
\barr{lll}
\nb^\mu\nb_\mu F_{\nu \la}& + &\l\{ R_{\rho \mu \nu \la}
F^{\rho\mu} + R^\rho_{\;\;\la} F_{\nu\rho} - R^\rho_{\;\;\nu} F_{\la \rho} \r\}
\\[8pt]
&+&\l[ \nb_\nu X_\la - \nb_\la X_\nu \r] = 0
\earr
\ee
In the derivations of [4-12] the terms in the curly bracket do
not appear.  This is because in the derivations, the eikonal
approximation is made in the Maxwell's equation and therefore
the Riemann and Ricci terms which arise on commuting the
covariant derivatives do not appear. In (43) however we see that
in the range of validity of the effective action (40), $\o^2
< \alpha / m^2_e$, the Riemann and Ricci terms in the
curly brackets of (48) which are present in Einstein's gravity
are larger than the terms in the square brackets of (48) which
arise from the loop corrections.  The photon velocities are
therefore given by the expression (26) for blackholes, and (39)
for Friedmann-Robertson-Walker metric, and is subluminal even in the
presence of higher derivative coupling terms.

{\it Acknowledgements:} We thank the anonymous referee for his  useful
suggestions.

\newpage
\noindent
\un{References}:
\begin{enumerate}
\item A.S.Eddington , {\it The Mathematical Theory of Relativity},
Cambridge University Press, (1957), p 176.
\item T.W. Noonan, {\it Ap. J.} {\bf 341} (1989) 786.
\item S.W.Hawking ,{Nature} {\bf 248} ,( 1974),  30.
\item I.T. Drummond and S.J. Hathrell, {\it Phy. Rev.} {\bf D22}
(1980) 343.
\item R.D. Daniels and G.M. Shore, {\it Nucl. Phys.} {\bf B425}
(1994) 634.
\item R.D. Daniels and G.M. Shore, {\it Phys. Lett.} {\bf B367}
(1996) 75.
\item J.L. Lattorre, P. Pascaul and R. Tarrach, {\it Nucl.
Phys.} {\bf B460} (1996) 379.
\item G.M. Shore, {\it Nucl. Phys.} {\bf B460} (1996) 379.
\item R. Lafrance and R.C. Myers, {\it Phys. Rev.} {\bf D51}
(1995) 2584.
\item I.B. Khriplovich, {\it Phys. Lett.} {\bf B346} (1995) 251.
\item A.D. Dolgov and I.B. Khriplovich, {\it Sov. Phy. JETP}
{\bf 58} (1983) 671.
\item P.F. Mende, in {\it String Quantum Gravity and Physics at
the Planck Scale}, Proceedigs of Erice Workshop, 1992, Ed. N.
Sanchez (World Scientific, Singapore, 1993).
\item Ya. B Zeldovich and I.D.Novikov , {\it Stars and
Relativity, Vol 1}, pg 197, Chicago University Press, Chicago,
(1971).  

\end{enumerate}
\end{document}